\documentclass[12pt]{article}
\usepackage{epsfig}
\usepackage{graphicx}
\usepackage[verbose]{wrapfig}
\let\section=\subsection     \let\subsection=\subsubsection

\newcommand{\bc} {\begin{center}}
\newcommand{\ec} {\end{center}}
\newcommand{\bqa}{\begin{eqnarray}}
\newcommand{\eqa}{\end{eqnarray}}
\newcommand{\nn}{\nonumber}

\newcommand{\ga}{\gamma}
\newcommand{\de}{\delta}

\newcommand{\si}{\sigma}
\newcommand{\om}{\omega}
\newcommand{\ta}{\tau}

\begin{document}
\hfill DESY 02-019\\[-3mm]
\begin{center}
   {\large \bf THERMAL DILEPTON RATES AND MESON}\\[2mm]
   {\large \bf SPECTRAL FUNCTIONS FROM LATTICE QCD}\\[5mm]
   I.~WETZORKE\\[1mm]
   {\small \it NIC/DESY Zeuthen, Platanenallee 6, D-15738 Zeuthen,
 Germany \\[5mm] }
\end{center}

\begin{abstract}\noindent
  Pseudo-scalar and vector meson correlation functions were calculated at
  temperatures below and above the deconfinement transition using
  ${\cal O}(a)$ improved Wilson fermions in quenched lattice QCD.
  The spectral functions were reconstructed from the correlator given only
  at discrete points in Euclidean time by means of the Maximum Entropy Method
  without a priori assumptions on the spectral shape. The deviation of the
  spectral functions from the one of a freely propagating quark anti-quark
  pair is examined above the critical temperature. At 1.5 and 3 $T_c$ the
  vector spectral function yields an enhancement of the dilepton rate over
  the Born rate in the energy interval $4 < \omega/T < 8$ and a sharp drop
  at energies below 2-3 T.
\vspace*{-2mm}
\end{abstract}

\section{Introduction}
The thermal dilepton spectrum is accessible in heavy ion collisions and
provides a key observable to study thermal properties of the medium at high
density and temperature \cite{Kapusta,Toimela_Alam}.
Thermally induced changes of the dilepton spectrum at low energies \cite{Abreu}
are expected to be influenced by non-perturbative in-medium modifications of
the quark anti-quark interactions. Furthermore such effects are directly
related to changes of the spectral shape in the vector meson channel.
In two-flavor QCD the differential dilepton rate is connected to vector
spectral function $\si_V(\om,\vec{p},T)$ through the relation
\bqa
\frac{dN_{l\bar l}}{d^4xd^4p} \;\;\equiv\;\;
\frac {dW} {d\omega d^3p} &=& 
\frac {5\alpha^2} {27\pi^2} \;
\frac {1} {\omega^2 \left({ e^{\om/T} -1 }\right)} \;
\sigma_V(\omega,\vec{p},T) \;.
\label{rate}
\eqa
At the same time the spectral densities are related to the Euclidean
correlation functions of hadronic currents
$J_H (\tau, \vec{x}) =
\bar{\psi}(\tau, \vec{x})\;\Gamma_H\;\psi(\tau, \vec{x})$
with $\Gamma_H \in [1,\ga_5,\ga_\mu,\dots]$ 
through the integral equation
\bqa
G_H(\tau, \vec{p},T) &=& \int {\rm d}^3 x \; e^{i\vec{p}\cdot\vec{x}}
\langle J_H(\tau, \vec{x}) J_H^\dag(0,\vec{0})\rangle\nn\\
&=& \int_0^\infty {\rm d} \om \;
\si_H(\om,\vec{p},T)\;
{ {\rm ch} (\om(\ta -1/2T)) \over {\rm sh} (\om/2T)} \;.
\label{corr}
\eqa
Although hadronic correlators can be calculated numerically in the 
framework of lattice regularized QCD for discrete Euclidean times
$\ta \in [0,1/T]$, the inversion of the integral equation still remains an
ill-posed problem. This situation improved considerably after the
implementation of the Maximum Entropy Method (MEM) for lattice QCD
\cite{Hatsuda}, which allows to obtain the most probable spectral function
without a priori assumptions on the spectral shape.

Most of the previous applications of MEM concentrated on the analysis of
hadronic properties like masses and decay width at zero temperature, where
the pole masses of the ground \cite{Hatsuda,Wetzorke} and even excited states
\cite{Yamazaki} could be obtained with quite satisfactory precision. The great
challenge, however, is to apply the method at non-zero temperature, where very
little about the spectral shape is known so far. 

\section{Thermal Meson Spectral Functions}
At finite temperature excited states and continuum-like contributions are
expected to gain in influence on the spectral shape. Thus it is mandatory to
verify that the method allows to reconstruct such spectral functions correctly.
A first step in this direction was the reconstruction of the spectral shape for
free massless quarks in the (pseudo-)scalar channel
$\si_{PS}^{free}=3/8 \; \pi^{-2} \om^2 \tanh(\om/4T)$, known from leading order
perturbation theory. In this case the relation (\ref{corr}) was used to obtain
discrete values of the correlator in Euclidean time \cite{Wetzorke}. On the 
other hand such a correlator is directly calculable on the lattice \cite{free},
which is shown in figure 1(a) for three different lattice
sizes. At small time separations the finite lattice cut-off effects are visible
compared to the free continuum correlation function indicated as solid
line. Nevertheless, the spectral shape could be reconstructed almost perfectly
already for temporal extents $N_\ta=8 \dots 16$ (see fig.~1(b)) by
introducing a lattice version of the integration kernel in equation
(\ref{corr}) for vanishing momentum
\bqa
G_H(\tau, T) =  \int_{0}^{\infty} {\rm d} \omega\;
\sigma_V (\omega,T)\; K_{L}(\tau,\omega, N_\tau)~~, 
\eqa
where $K_{L}(\tau,\omega, N_\tau)$ is the finite lattice approximation 
of the continuum kernel
\bqa
K_{L} (\tau,\omega,N_\tau) 
= {2\omega \over T} \sum_{n=0}^{N_\tau -1}
{\exp (- i 2\pi n\tau T) \over 
(2 N_\tau \sin(n\pi /N_\tau ))^2 + (\omega/T)^2} 
\quad.
\eqa
\begin{figure*}
\epsfig{file=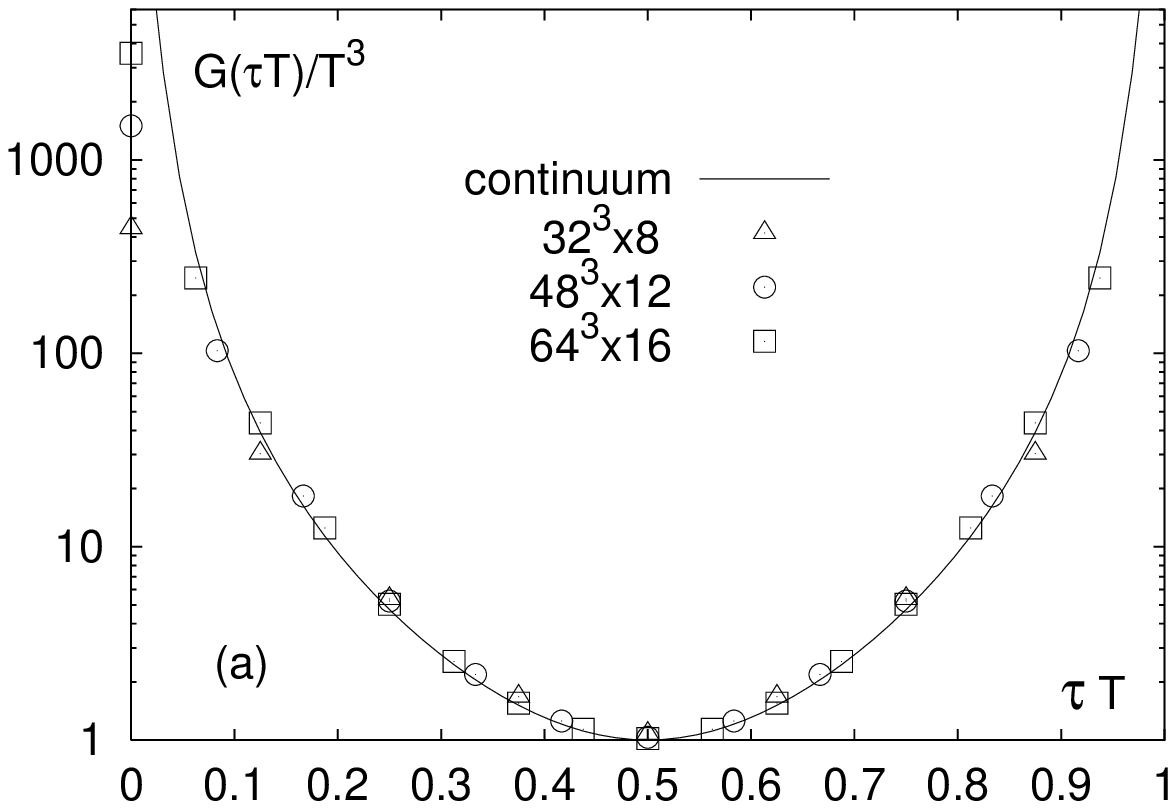,width=70.0mm}
\epsfig{file=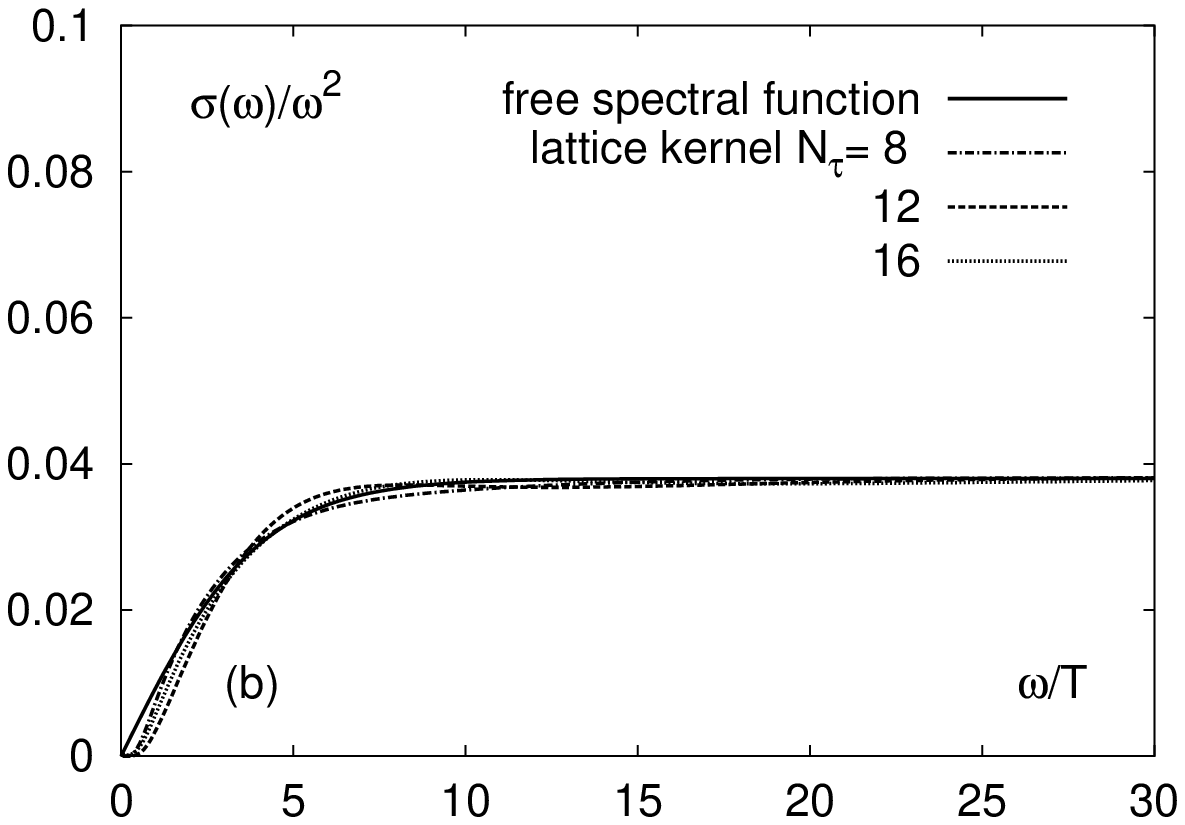,width=70.0mm}
\vskip-3mm
\label{free_lat}
\caption{Free scalar correlators (a) and reconstructed spectral
  shape (b)}
\end{figure*}
Since the applicability of MEM at finite temperature has been verified, one can
proceed to analyze meson spectral functions below and above the critical
temperature of the deconfinement transition. We have calculated pseudo-scalar
and vector meson correlation functions on large isotropic lattices of sizes
\mbox{$(24-64)^3 \times 16$} in a temperature range 0.4 to 3.0 $T_c$ using
${\cal O}(a)$ improved Wilson fermions \cite{clover} in quenched
lattice QCD. Below $T_c$ up to five quark mass values were chosen in order to
obtain a reliable extrapolation to the chiral limit ($m_q \to 0$). Above the
critical temperature the simulations could be performed directly at
approximately zero quark mass, since there are no longer massless Goldstone
modes present \cite{Dilept}.

Compared to the sharp $\de$-function like peaks observed in the spectral
functions at $T=0$ \cite{Yamazaki} a broadening of the peaks corresponding to 
particle poles and a reduction in height can be observed at finite
temperature. Figure 2 shows an example at 0.4 $T_c$, where the error bars
indicate the uncertainty in the given energy region calculated from the
covariance matrix of the spectral
\begin{wrapfigure}[13]{o}{71mm}
\vskip-1mm
\epsfig{file=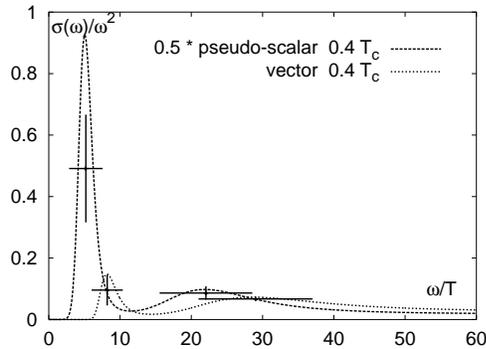,width=70mm}
\vskip-3mm
\label{0.4Tc}
\caption{Meson spectral functions for a ratio of $m_\pi/m_\rho$ = 0.6 at 0.4
  $T_c$}
\end{wrapfigure}
function \cite{Bryan}. The ground state
masses extracted from the spectral functions at 0.4 $T_c$ agree very well with
zero temperature data \cite{Goeckeler}. Increasing the temperature towards
$T_c$ the vector meson spectral function is investigated at fixed pion
mass. The ground state peak experiences a further broadening and is slightly
shifted towards higher energies. It remains to be clarified to what extent this
effect is induced by the restricted temporal dimension and limited statistics.

Above the critical temperature only a gradual approach towards the free
correlation function is observed in the (pseudo-)scalar channel, while the
vector meson correlator is much closer to the free quark behavior already at
1.5 $T_c$ (see fig.~3(a)). Analyzing directly the ratio of the
correlation function in the vector channel to the corresponding free curve
calculated for the same lattice size yields a deviation of about 10 \%. The
free correlator is approached from above, while simple quasi-particle
pictures lead to a reduction of $G_V(\ta,T)$ relative to $G^{free}_V(\ta,T)$
\cite{Kapusta,Mustafa}.
The deviations from the free quark behavior are also evident from the reconstructed
spectral functions in figure 3(b). While the second enhancement
over the free curve visible in fig. 2 and 3(b) at
energies larger than $\om/T = 16$ is presumably a lattice artifact due to the
heavy fermion doublers of the Wilson fermion formulation \cite{Yamazaki}, the
peak at about $\om/T = 6$ shows the actual physical effect of persisting
interactions between quarks and gluons in both channels up to 3 $T_c$. Compared
to the zero temperature results \cite{Hatsuda,Wetzorke,Yamazaki} the peaks are
rather broad and less pronounced in height, but might sharpen by
increasing the statistics and the lattice size \cite{Hatsuda}. 

\begin{figure*}
\epsfig{file=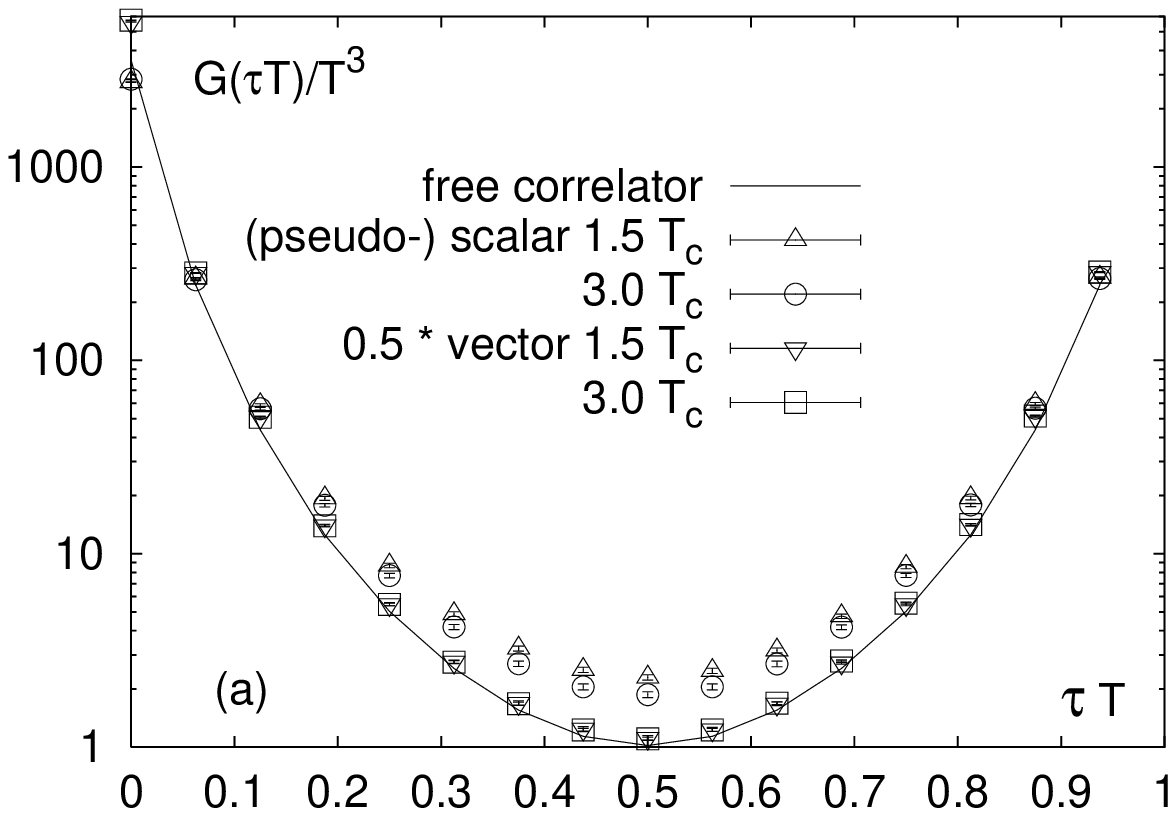,width=70.0mm}
\epsfig{file=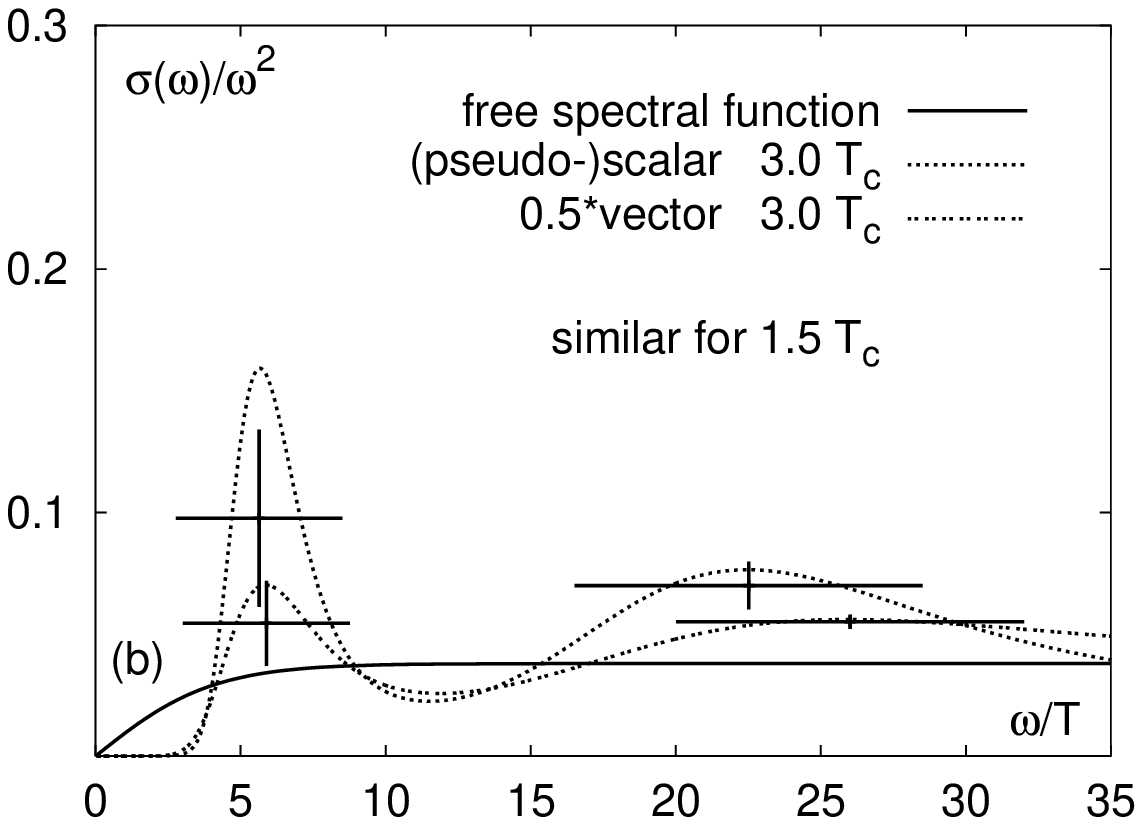,width=70.0mm}
\vskip-3mm
\label{above_Tc}
\caption{Meson correlators (a) and reconstructed spectral functions
  (b)}
\end{figure*}
\section{Thermal Dilepton Rates}
Once the most probable vector meson spectral function is reconstructed from the
correlation function in Euclidean times by means of the Maximum Entropy Method,
the remaining step to obtain the differential dilepton rate is straight
forward. The spectral functions shown in figure 4(a) were obtained
from correlators on the largest lattice ($64^3 \times 16$) with approximately
zero quark mass at temperatures 1.5 and 3.0 $T_c$ \cite{Dilept}. In addition
to the purely statistical error-band obtained from a Jackknife analysis, the
uncertainty of the uniqueness of the result incorporated in MEM is shown in the
insertion as error bars for four energy intervals. This leads to the conclusion
that the data sample is large enough to yield statistically significant
results.

\begin{figure*}
\vspace*{0.4cm}
\hspace*{-0.6cm}\epsfig{file=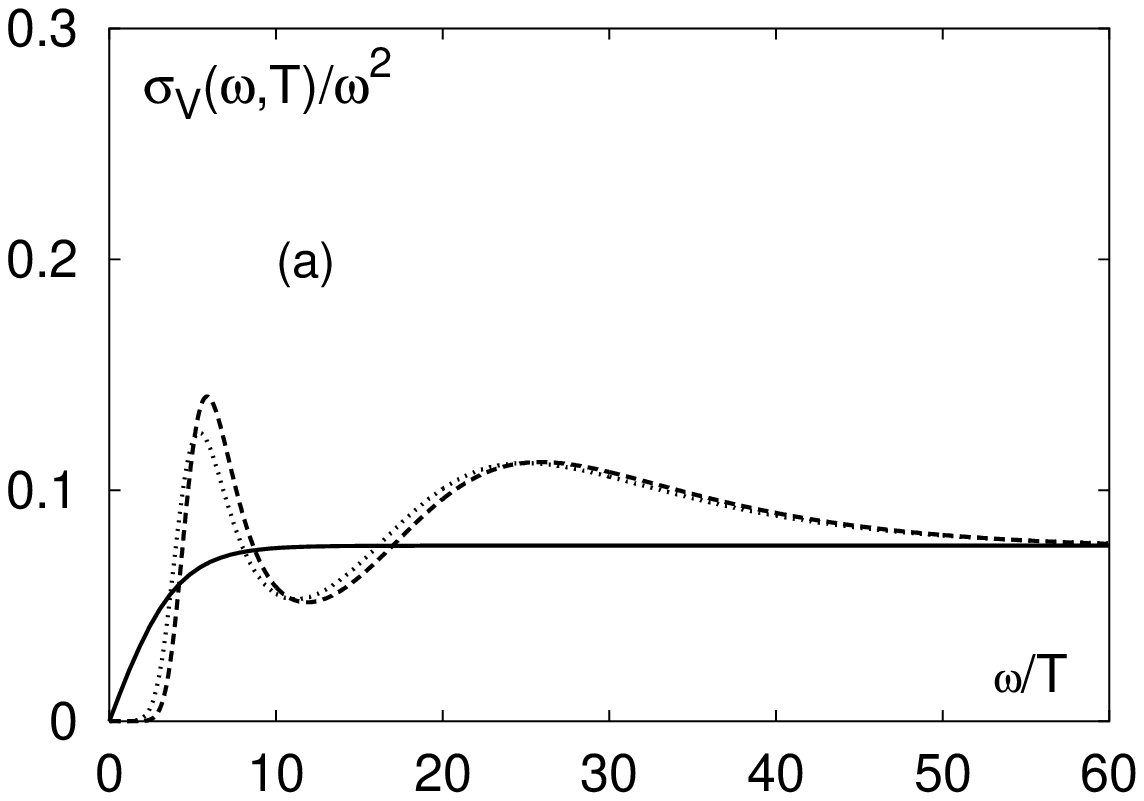,width=75mm}
\hspace*{-0.4cm}\epsfig{file=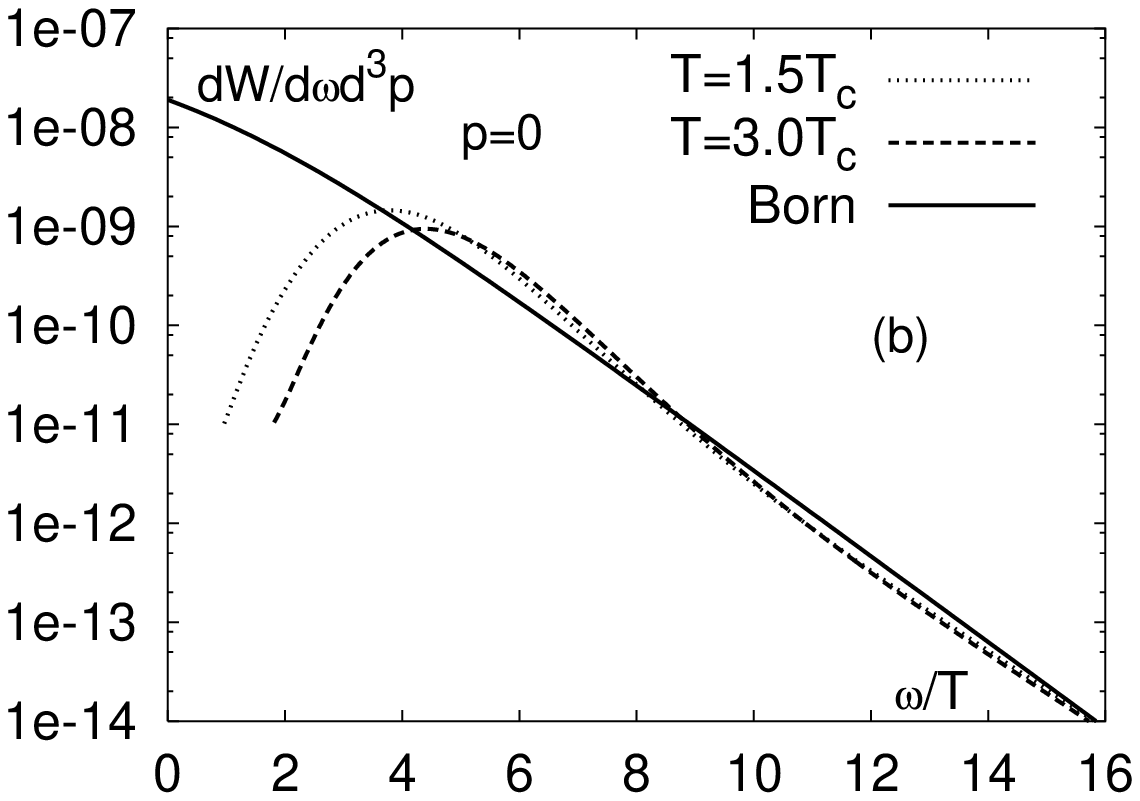,width=75mm}
\setlength{\unitlength}{1cm}
\begin{picture}(0.1,0.1)
\put(2.82,2.8){\epsfig{file=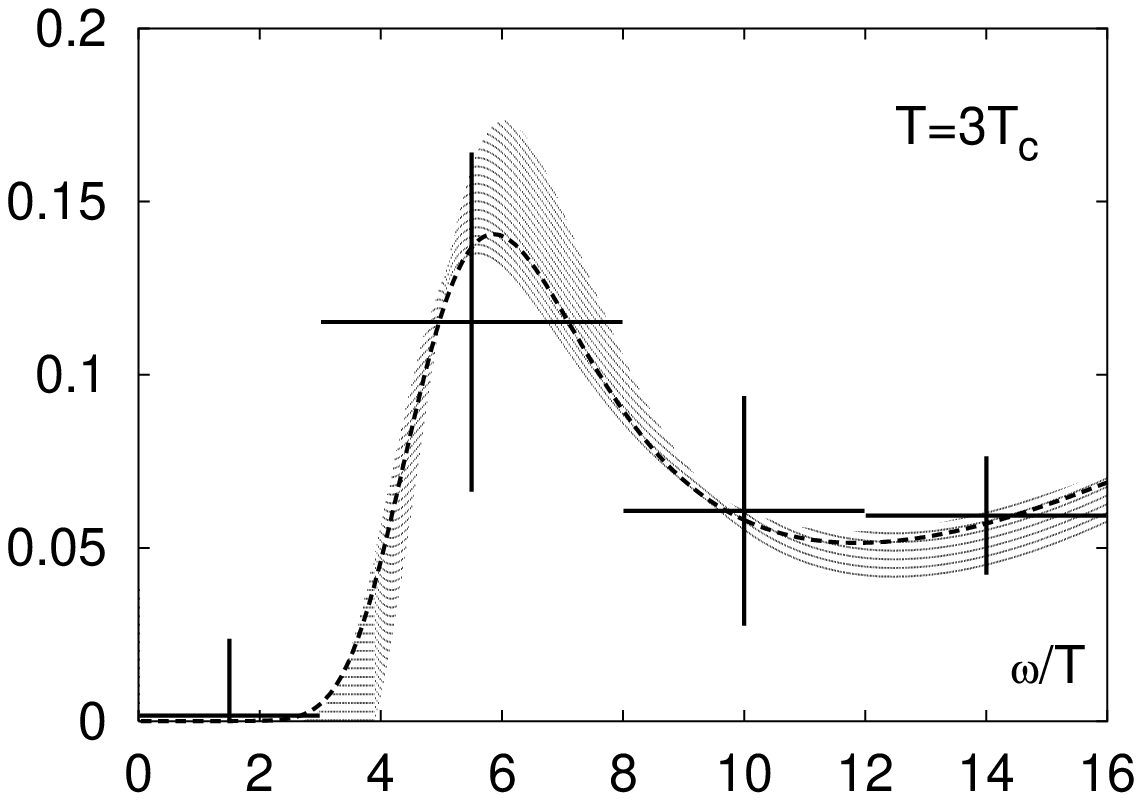,width=38mm}}
\end{picture}
\vskip-8mm
\label{dilep}
\caption{Vector spectral function (a) and differential dilepton rate (b)}
\end{figure*}
The differential dilepton rate can now be calculated in employing relation
(\ref{rate}) projected onto zero momentum. Figure 4(b) illustrates
the corresponding dilepton rates for both temperatures. In the high temperature
limit the leading order perturbative result can be obtained by using free
massless quark propagators in the calculation of the current-current correlation
functions. For vanishing momentum the spectral function in the vector channel
is given by $\si_V^{free}=3/4 \; \pi^{-2} \om^2 \tanh(\om/4T)$ , which leads to
the reduction of the differential dilepton rate to the Born rate
\bqa
{{\rm d} W^{\rm Born} \over {\rm d}\omega {\rm d}^3p}(\vec{p}=0) = 
{5 \alpha^2 \over 36 \pi^4} {1\over ({\rm e}^{\omega/2T} + 1)^2}
\quad .  
\eqa
The broad peak visible in the spectral functions at around $\om/T=5-6$
results in an enhancement of the respective dilepton rates in the region
$\om/T=4-8$ compared to the Born rate, which seems to scale with the
temperature. For even larger energies the close agreement of the
differential dilepton spectrum with the Born rate is evident. 

The most striking feature of the observed dilepton rate is certainly the sharp
drop at energies below $\om/T=2-3$. This effect is in contrast to hard thermal
loop resummed perturbation theory \cite{htl} as well as to 2-loop perturbative
calculations \cite{Aurenche}, which lead to a divergent vector spectral
function in the limit $\om \to 0$. If this suppressions persists in future
investigations closer to the critical temperature, there would be no thermal
contribution the dilepton rate at small energies during the expansion of the
hot medium created in heavy ion collisions. This is consistent with the present
results at SPS energies \cite{ceres} and might be observable in the dilepton
spectra at RHIC or LHC energies. 

\section*{Acknowledgements}
The work summarized in this talk is part of a very lively collaboration with
F.~Karsch, E.~Laermann, P.~Petreczky and S.~Stickan. Furthermore, I want to
thank the organizers of the Hirschegg Workshop on Ultrarelativistic Heavy-Ion
Collisions for the opportunity to present this contribution and many
of the participants for fruitful discussions.

\end{document}